# The TES-based Cryogenic AntiCoincidence Detector of ATHENA X-IFU:
# Validation of the thermal end-to-end simulator towards the updated Demonstration Model (DM 1.1)

Matteo D'Andrea, Claudio Macculi, Simone Lotti, Luigi Piro, Andrea Argan, Gabriele Minervini, Guido Torrioli, Fabio Chiarello, Lorenzo Ferrari Barusso, Edvige Celasco, Flavio Gatti, Daniele Grosso, Manuela Rigano, Daniele Brienza, Elisabetta Cavazzuti, Angela Volpe

*Abstract*— The Cryogenic AntiCoincidence Detector (CryoAC) is a key element of the X-ray Integral Field Unit (X-IFU) on board the future ATHENA X-ray observatory. It is a TES-based detector designed to reduce the particle background of the instrument, thereby increasing its sensitivity. The detector design is driven by an end-to-end simulator which includes the electro-thermal modelling of the detector and the dynamics of its readout chain. Here, we present the measurements carried out on the last CryoAC single pixel prototype, namely DM127, in order to evaluate the critical thermal parameters of the detector and consequently to tune and validate the CryoAC end-to-end simulator.

## I. INTRODUCTION

ATHENA is the next ESA flagship X-ray observatory (0.2 - 12 keV) [1]. The mission has recently undergone a successful redefinition process to fit within new ESA constraints and is now scheduled for launch in 2037 [2]. The X-IFU is one of the two instruments on the payload. It is a cryogenic spectrometer based on a large array of ~1500 TES microcalorimeters, operating with a thermal bath at 50 mK and achieving outstanding spectral resolution (~3 eV at 7 keV) [2][3].

The X-IFU needs to host a Cryogenic AntiCoincidence

detector (CryoAC [4]) as the TES array alone is not able to discriminate between target X-ray photons and background particles. This is a second TES-based detector, placed < 1 mm below the TES array. While X-rays are absorbed in the TES array, background particles deposit energy in both detectors, producing a coincidence signal that allows these unwanted events to be vetoed. The CryoAC reduces the X-IFU particle background by a factor of ~ 30, increasing its sensitivity and enabling a significant part of the mission science [5].

The CryoAC is a kind of instrument-inside-the-instrument, with independent electronics and a dedicated data processing chain. It is a 4-pixel detector, based on large area silicon absorbers (~1 cm² each one) sensed by Ir/Au TES. Each pixel is readout by a DC-SQUID and features an on-chip heater for calibration purposes. The detector is required to have a wide energy bandwidth, with a low energy threshold < 20 keV and a saturation energy > 1 MeV. This allows to sense the energy losses of background particles in the absorber, distributed around the most-probable MIP (Minimum Ionizing Particle) deposition at ~ 150 keV. Furthermore, it shall have high efficiency (< 0.014% of missed particles), low dead time (< 1%) and good time tagging accuracy (10 µs at 1σ).

Recently the CryoAC baseline has moved from an athermal design [6], developed to optimize the athermal phonon collection and speed up the pulse response, to a more classical microcalorimeter thermal scheme. This provides a simpler and more robust baseline for the detector development, while maintaining efficiency and timing performance within the requirements [7]. This change in baseline will result in an updated version of the CryoAC Demonstration Model, i.e. the CryoAC DM 1.1, which will be tested with the NASA TES array detector and its Time Division Multiplexing (TDM) readout inside the DM1.1 Focal Plane Assembly (FPA) developed by SRON [8].

The detailed design of the detector is now being driven by a new end-to-end CryoAC thermal simulator. Here we present the measurements carried out to tune and validate this simulator using the latest single-pixel CryoAC thermal prototype.

This work has been supported by ASI (Italian Space Agency) contracts n. 2019-27-HH.0, 2019-27-HH.1-2021, 2019-27-HH.2-2023.
*Corresponding author: Matteo D'Andrea (email: matteo.dandrea@inaf.it, ORCID: https://orcid.org/0000-0002-5139-4578).*

Matteo D'Andrea, Claudio Macculi, Simone Lotti and Luigi Piro are with INAF/IAPS, Via del Fosso del Cavaliere 100, 00133 Rome, Italy.
Andrea Argan and Gabriele Minervini are with INAF Headquarters, Viale del Parco dei Mellini 84, 00136, Rome, Italy.
Guido Torrioli and Fabio Chiarello are with CNR/IFN Roma, Via del Fosso del Cavaliere 100, 00133, Rome, Italy.
Lorenzo Ferrari Barusso, Edvige Celasco, Flavio Gatti, Daniele Grosso and Manuela Rigano are with University of Genoa (Phys Dept.), Via Dodecaneso 33, 16146 Genoa, Italy.and with INFN Genova, Via Dodecaneso 33, 16146 Genoa, Italy.
Daniele Brienza, Elisabetta Cavazzuti and Angela Volpe are with ASI, Via del Politecnico snc, 00133, Rome, Italy.

## II. THE CRYOAC THERMAL SIMULATOR

We have developed the CryoAC end-to-end simulator to drive the detector design and conduct its performance assessment. Its core is the CryoAC thermal simulator, which solves time-by-time the equations describing the electro-thermal state of the system and the dynamics of its readout chain, based on a DC-SQUID operated in a Flux Locked Loop (FLL) scheme. A full description of the simulator and its first application to evaluate the performance of the CryoAC is reported in [7]. The thermal model of the detector used in the simulator is shown in Fig.1.

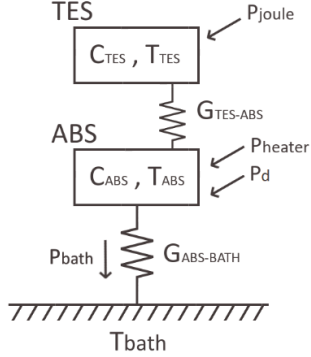

**Fig. 1.** Thermal circuit representing a CryoAC pixel. $C_{TES}$ and $T_{TES}$ are the TES electrons heat capacity and temperature. Pjoule is the joule power dissipated by the TES bias. The TES is connected to the silicon absorber through the thermal conductance $G_{TES-ABS}$ (total equivalent conductance including TES electron-phonon conductance and the Kapitza interface conductance). $C_{ABS}$ and $T_{ABS}$ are the absorber phonons heat capacity and temperature. Pheater is the power dissipated on the absorber by the on-chip heater. Pd is the power dissipated on the absorber by incoming particles/photons. $G_{ABS-BATH}$ is the thermal conductance between the absorber and the thermal bath at Tbath temperature, through which the total power Pbath = Pjoule + Pheater + Pd flows.

## III. THE CRYOAC PROTOTYPE DM127

The simulator has been tuned and validated using the latest single-pixel CryoAC thermal prototype, namely DM127 (Fig. 2). It is a single-pixel DM-like sample based on a suspended 1 cm² silicon absorber, sensed by a single Ir/Au TES. It is read-out by a DC-SQUID produced by VTT [9] (model M4A, ret M) and operated by a commercial Magnicon XXF-1 FLL electronics. The TES is connected in parallel with a shunt resistor $R_S = 8$ mΩ. The detector also features an on-chip heater for diagnostic and calibration purposes ($R_H = 394$ Ω, measured at 4-wires at 50 mK). The detailed description of the sample and the report of its first test campaign are given in [10].

The TES transition of the sample presents a problem due to a fluorides contamination in the reactive ion etching of the TES, resulting in a broad, poorly repeatable transition curve (Fig. 3). Despite this issue, it was possible to properly operate the sample around T ~ 100 mK, and exploit its broad transition to measure the thermal parameters of the detector over a wide temperature range.

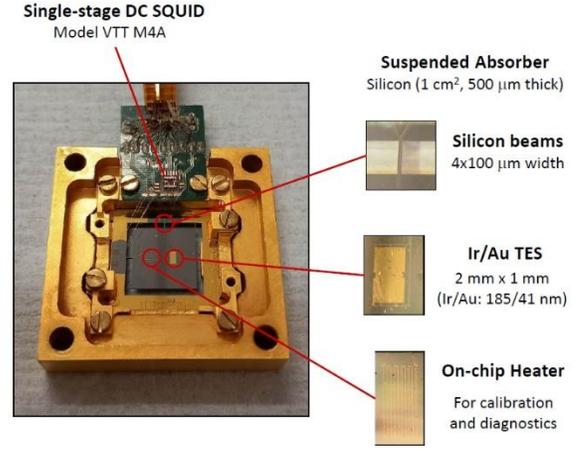

**Fig. 2.** CryoAC DM-like thermal prototype, namely DM127.

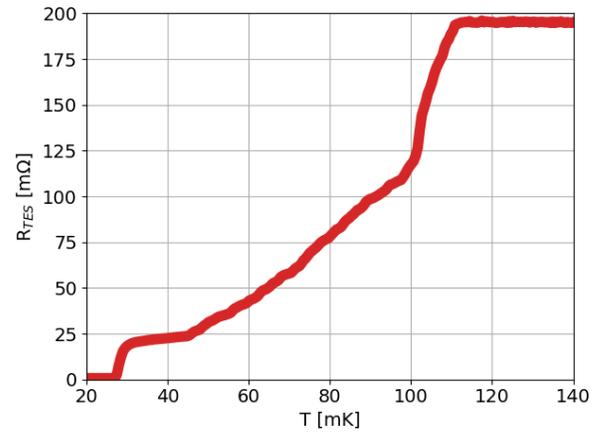

**Fig. 3.** CryoAC DM 127 Transition curve.

## IV. DETECTOR THERMAL PARAMETERS MEASUREMENTS

We have exploited the broad DM127 TES transition and the on-chip heater to perform novel, accurate measurements of the detector thermal parameters, which are reported in the following subsections.

### A. Absorber to bath thermal conductance ($G_{ABS-BATH}$)

The thermal conductance between the absorber and the thermal bath is given by the narrow silicon bridges that define the suspended structure of the CryoAC absorber. We have measured it by exploiting the DM127 on-chip heater.

Using the heater, we have measured the $R_{TES}$ vs $P_{Heater}$ curve (i.e. the TES transition induced by the power dissipated by the heater on the absorber) at different thermal bath temperatures (Fig. 4). The TES resistance is measured at each point by injecting a small current into the TES bias circuit ($I_{BIAS} < 10$ μA), so the power dissipated across the TES itself is negligible ($P_{TES} < 0.2$ pW). Therefore, in this measurement we can assume that the TES and the absorber are at the same temperature, which can be extracted by the measured TES transition curve in Fig. 3:

$$T_{ABS} = T_{TES} = f(R_{TES}) \qquad (1)$$

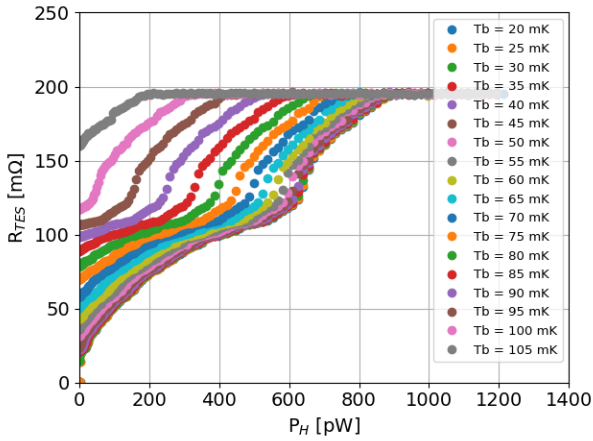

**Fig. 4.** Measured $R_{TES}$ vs $P_{Heater}$ curves at different thermal bath temperatures.

In Fig. 4, we can select all the points corresponding to a certain resistance value, which corresponds to a certain TES and absorber temperature, and plot the heater power required to reach this resistance as a function of the thermal bath temperature (Fig. 5). The data points can then be fitted using the formula:

$$P_H = k_{AB}(T_{ABS}^{n_{AB}} - T_B^{n_{AB}}) \qquad (2)$$

The $k_{AB}$ and $n_{AB}$ parameters define the thermal conductance of the link [11]:

$$G_{ABS-BATH}(T) = n_{AB} \cdot k_{AB} \cdot T^{(n_{AB}-1)} \qquad (3)$$

Note that the fitted $n_{AB}$ parameter ($n_{AB} = 4.03 \pm 0.03$) is well compatible with the value n=4, as expected for a thermal coupling given by phonons in a crystal [11].

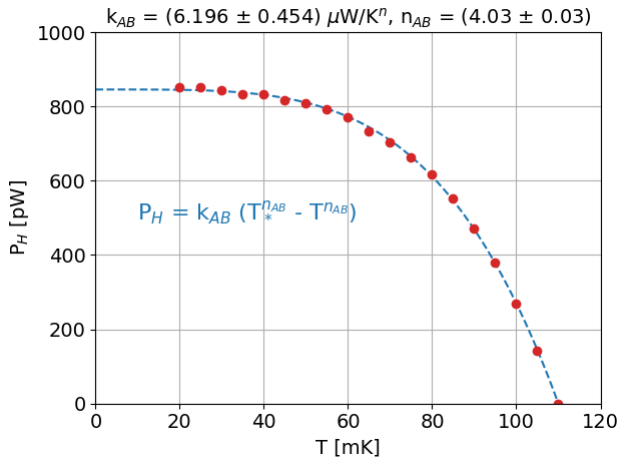

**Fig. 5.** Power injected on the absorber to bring the TES to a temperature of $T_* = 110$ mK, as a function of the thermal bath.

Once the $n_{AB}$ parameter has been assessed, we can measure $k_{AB}$ with higher precision than the result shown in Fig.5. In fact, we can use all the points of the $R_{TES}$ vs $P_{Heater}$ curves to generate the plot in Fig. 6, where the heater power is plotted directly versus $\left(T_{TES}^4 - T_B^4\right) = \left(T_{ABS}^4 - T_B^4\right)$, with the temperature data evaluated by inverting the measured transition curve. Finally, the data points can be fitted by a linear regression, in order to obtain $k_{AB}$ (eq. (2)).

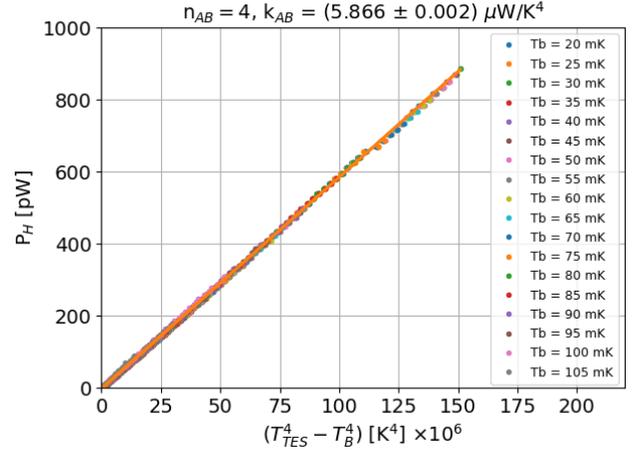

**Fig. 6.** Final assessment of the thermal conductance between the absorber and the thermal bath.

The final assessment of the parameters describing the absorber to bath thermal conductance is therefore:

$$n_{AB} = 4, \, k_{AB} = (5.866 \pm 0.002) \, \mu W/K^4 \qquad (4)$$

Note that this value is fully compatible with the old, less accurate assessment obtained by measuring the CryoAC DM 1.0 in 2019 [12].

### B. TES to absorber thermal conductance ($G_{TES-ABS}$)

The thermal conductance from the TES to the absorber, mainly due to the electron-phonon decoupling in the TES, is a particularly critical parameter affecting the detector dynamics. To measure it, we have first extracted the $R_{TES}$ vs $P_{TES}$ curve (i.e. the TES transition induced by dissipating joule power on the TES) from the characteristics I-V curves measured for the sample at different thermal bath temperatures (Fig. 7).

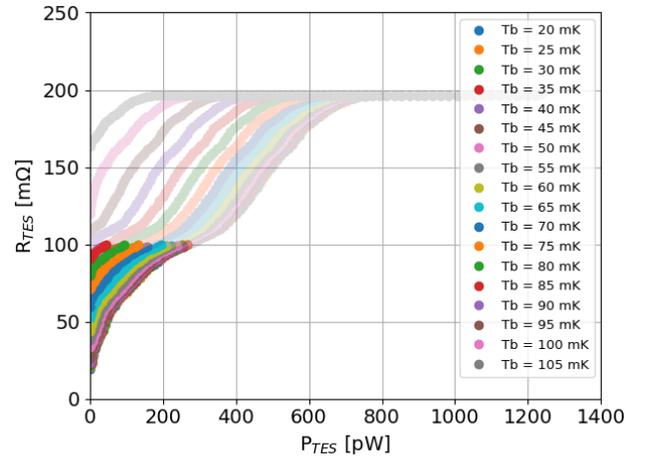

**Fig. 7.** Measured $R_{TES}$ vs $P_{TES}$ curves at different thermal bath temperatures, with the points at $R_{TES} < 100$ m$\Omega$ highlighted.

We have then selected from these curves the region at $R_{TES} < 100$ m$\Omega$, where the TES transition has a low slope (see Fig. 3) and thus the TES current-responsivity is likely to be negligible, so we can assume $R_{TES}(T_{TES}, I_{TES}) \sim R_{TES}(T_{TES}, I_{TES} = 0)$. There, for each point we have evaluated the temperature of the TES from the transition curve (which has been measured in the limit $I_{TES} \sim 0$), and the temperature of the absorber by using the absorber to bath conductance parameters previously measured:

$$T_{TES} = f(R_{TES}) \tag{5}$$

$$T_{ABS} = (P_{TES}/k_{AB} + T_B{}^{n_{AB}})^{(1/n_{AB})} \tag{6}$$

Note that eq. (6) is valid since in this measurement the joule power dissipated on the TES is also the only one flowing to the thermal bath ($P_{HEATER} = 0$ and $P_D = 0$). Finally, we have fitted the data with the equation:

$$P_{TES} = k_{TA}(T_{TES}{}^{n_{TA}} - T_{ABS}{}^{n_{TA}}) \tag{7}$$

The $k_{TA}$ and $n_{TA}$ parameters define the thermal conductance from the TES to the absorber:

$$G_{TES-ABS}(T) = n_{TA} \cdot k_{TA} \cdot T^{(n_{TA}-1)} \tag{8}$$

The fit is shown in Fig. 8. We have assumed $n_{TA} = 6$, which is the expected power index for a thermal link dominated by the electron-phonon decoupling [11].

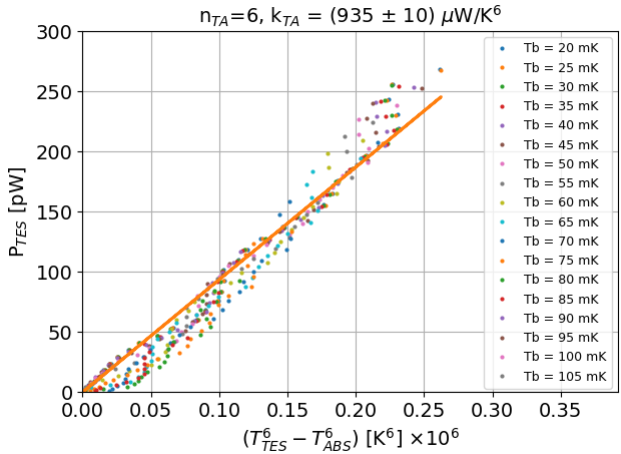

**Fig. 8.** Assessment of the thermal conductance between the TES and the absorber.

We have then repeated the analysis by varying the $n_{TA}$ parameter. The Sum of Squared Residual (SSR) of each fit is reported in Fig. 9, which shows that the optimal index for this dataset is indeed in the range $6 < n_{TA} < 7$.

Our final assessment of the parameters describing the TES to absorber thermal conductance is:

$$n_{TA} = 6, k_{TA} = (935 \pm 10) \text{ μW}/K^6 \tag{9}$$

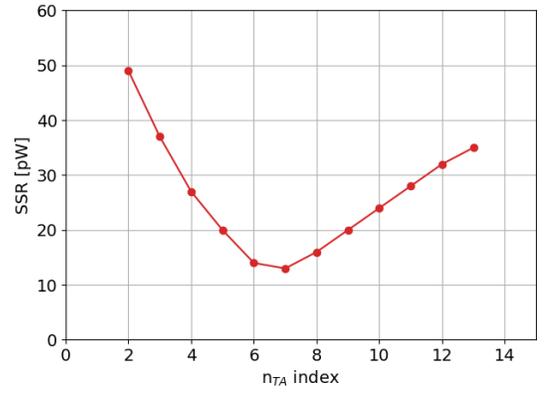

**Fig. 9.** Assessment of the parameter $n_{TA}$.

Note that the resulting thermal conductance per unit of TES volume ($V_{TES}$) is:

$$\boldsymbol{G_{TES-ABS}/V_{TES} = 4.0 \cdot 10^4 \ W/(K \ m^3)} \quad \text{at 114 mK} \quad (10)$$

which is close to the value reported in the literature for an Ir TES ($G_{TES-ABS}/V_{TES} = 5.4 \cdot 10^4 \ W/(K \ m^3)$ at 114 mK, from [13]).

### C. Absorber and TES heat capacity ($C_{ABS}$ and $C_{TES}$)

The heat capacities of a TES microcalorimeter are usually extracted from its complex impedance measurements [11]. Unfortunately, a setup to perform such measurements was not available, so the heat capacities of absorber and TES have been preliminary evaluated starting from theoretical values and corrected by arbitrary scaling correction factors in order to match the experimental data from DM127. The analysis is presented at the end of this section. It is worth of noting that such correction factors are independent of temperature, so they are a kind of scaling factor to be applied, probably depending on the real characteristics of the silicon and TES properties. For the purposes of the CryoAC simulator, this approach is acceptable. Indeed, the simulator is mainly used to test the CryoAC trigger algorithm and to evaluate the detector dead time [7], and in this context an accuracy of a several percent in reproducing pulse heights and shapes is sufficient.

The heat capacity of the silicon absorber is dominated by the contribution of the phonons in the crystal:

$$C_{ABS} = C_{ABS,phonons} \tag{11}$$

This is described in the simulator by the Debye theory [14]:

$$C_{ABS,phonons} = S_{ABS} \cdot 12/5 \cdot \pi^4 \cdot R \cdot \rho_{Si}/A_{Si} \cdot V \cdot T_{ABS}{}^3/\theta_{Si}^3 \tag{12}$$

where the relevant parameters are listed in Tab.1.

**Tab. 1.** Parameters used in C_ABS evaluation

| Parameter | Notes |
|-----------|-------|
| $S_{ABS}$ = 1.8 | Scaling correction factor |
| $R$ = 8.314472 J/(mol · K) | Universal gas constant |
| $\theta_{Si}$ = 645 K | Si Debye temperature |
| $V$ = 500 μm · 10 mm · 10 mm | Absorber volume |
| $\rho_{Si}$ = 2.329 g/cm³ | Si density |
| $A_{Si}$ = 28.085 g/mol | Si atomic weight |

The final value of the absorber heat capacity evaluated at different temperatures is reported in Tab. 2.

**Tab. 2.** Absorber heat capacity evaluated at different temperatures

| T [mK] | $C_{ABS}$ [pJ/K] |
|--------|-----------------|
| 20 | 0.4 |
| 50 | 6.8 |
| 75 | 22.8 |
| 100 | 54.1 |

The TES heat capacity is dominated by the contribution of the electrons in the bilayer:

$$C_{TES} = C_{TES,electrons} \quad (13)$$

This is described in the simulator according to the corresponding model for metals and the BCS theory [14]:

$$C_{TES,electrons} = S_{TES} \cdot ( 2.43 \cdot \rho_{Ir}/A_{Ir} \cdot V_{Ir} \cdot \gamma_{e,Ir} + \rho_{Au}/A_{Au} \cdot V_{Au} \cdot \gamma_{e,Au}) \cdot T_{TES} \quad (14)$$

where the relevant parameters are listed in Tab.3.

**Tab. 3.** Parameters used in C_TES evaluation

| Parameter | Notes |
|-----------|-------|
| $S_{TES}$ = 0.1 | Scaling correction factor |
| 2.43 | BCS factor |
| $\rho_{Ir}$ = 22.56 g/cm³ | Ir density |
| $A_{Ir}$ = 192.217 g/mol | Ir atomic weight |
| $\gamma_{e,Ir}$ = 3.20 · 10⁻³ J/(K² · mol) | Ir molar electronics specific heat |
| $V_{Ir}$ = 185 nm · 0.2 mm · 0.1 mm | Ir volume in the TES |
| $\rho_{Au}$ = 19.3 g/cm³ | Au density |
| $A_{Au}$ = 196.967 g/mol | Au atomic weight |
| $\gamma_{e,Au}$ = 0.729 · 10⁻³ J/(K² · mol) | Au molar electronics specific heat |
| $V_{Au}$ = 41 nm · 0.2 mm · 0.1 mm | AU volume in the TES |

The final value of the TES heat capacity evaluated at different temperatures is reported in Tab. 4.

**Tab. 4.** TES heat capacity evaluated at different temperatures

| T [mK] | $C_{TES}$ [pJ/K] |
|--------|-----------------|
| 20 | 0.7 |
| 50 | 1.7 |
| 75 | 2.6 |
| 100 | 3.4 |

The scaling correction factors S_ABS and S_TES have been tuned to match the amplitude and shape of the pulses acquired by the DM127. Fig. 10 shows the comparison between the simulator output (without the noise contributions) and the acquired average pulses at two different energies (5.9 keV from a ⁵⁵Fe source and 60 keV from a ²⁴¹Am source), assuming unitary scaling correction factors (i.e. no modification with respect to the theoretical capacity values: k_ABS = k_TES = 1). Note that the simulated pulses have a slightly smaller amplitude and a much slower rise time compared to the real data.

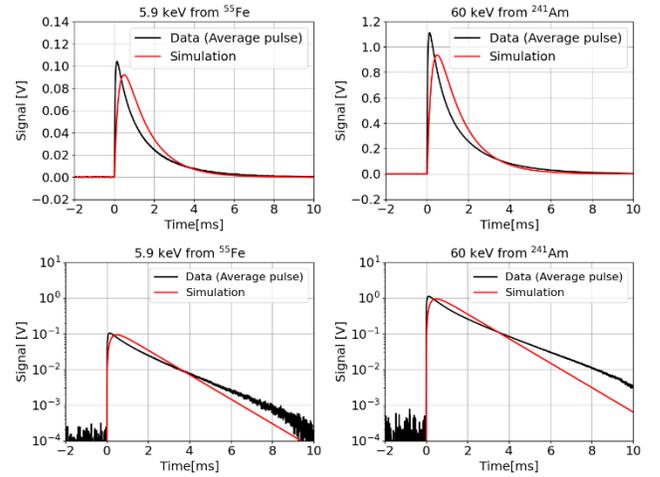

**Fig. 10.** Comparison between simulator output (red lines) and DM127 data (black lines), with unitary heat capacity scaling correction factors (S_ABS = 1 and S_TES = 1). (Top): linear scale plots, to compare pulse heights. (Bottom): log-scale plots, for better comparison of time constants.

Fig. 11 shows the same comparisons once considered the optimal scaling correction factors reported above (S_ABS = 1.8 and S_TES = 0.1). In this case there is a better agreement between the simulations and the real data, fully sufficient for the simulator purposes.

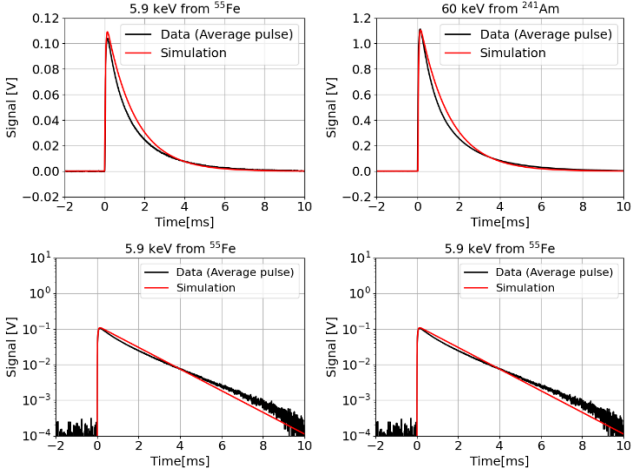

**Fig. 11.** Comparison between simulator output (red lines) and DM127 data (black lines) with the optimal heat capacity scaling correction factors ($S_{ABS} = 1.8$ and $S_{TES} = 0.1$). (Top): linear scale plots, to compare pulse heights. (Bottom): log-scale plots, for better comparison of time constants.

## V. Thermal Simulator Validation

Once the critical thermal parameters had been assessed, we have validated the simulator by comparing its output with real data acquired by DM127, using the thermal parameters extracted from the DM127 itself. We have already shown a comparison between data and simulations in Fig. 11, which shows a fine agreement in pulse shapes and pulse heights at two different energies. In the following subsections, we report the comparison in terms of detector noise, pulse heights up to the saturation regime and different detector operating points. The previously evaluated correction factors for the heat capacities are adopted. The DM127 has been operated at $R_0 = 139$ m$\Omega$, corresponding to $T_{TES,0} = 102$ mK.

### A. Detector Noise

The detector noise is due to several contributions, which are described in the simulator both at the cold stage and at the warm electronics level (see [7] for details). Fig. 12 shows a measured noise spectrum with the different theoretical contributions overplotted.

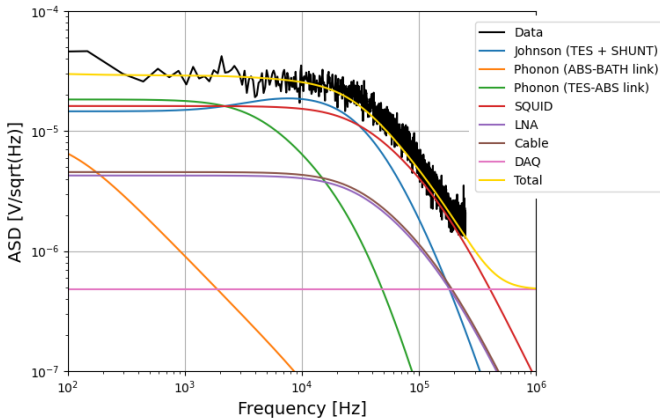

**Fig. 12.** DM127 noise spectrum (black line) with the theoretical contributions overplotted.

The comparison between the simulated and acquired detector noise spectra is then shown in Fig. 13, while Fig. 14 compares a simulated 5.9 keV pulse (including noise) with a real acquired pulse. The real data and the simulations are in good agreement. Quantitatively, the comparison of the RMS noise values is within a few percent, as reported in Tab. 5.

**Tab. 5.** Comparison of acquired and simulated RMS noise

|  | **Real data** | **Simulation** |
|---|---|---|
| **RMS noise (1σ)** | 5.46 mV | 5.20 mV |

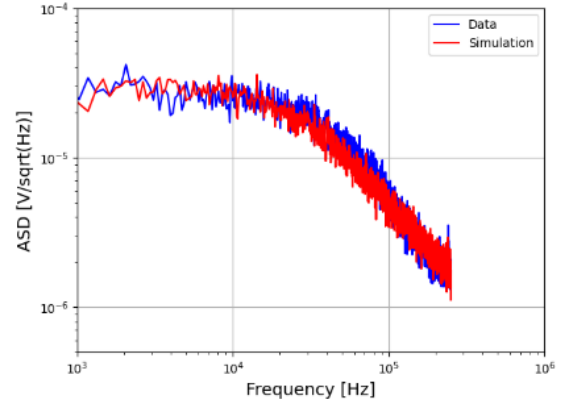

**Fig. 13.** Comparison between simulated noise spectrum (red line) and DM127 data (blue line).

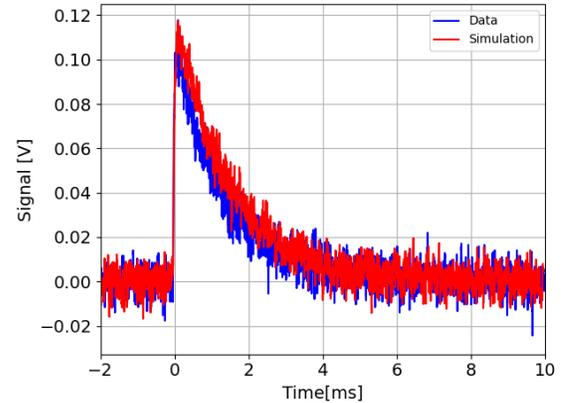

**Fig. 14.** Comparison between simulator output (5.9 keV pulse with noise, red lines) and corresponding DM127 data (blue line).

### B. Saturation regime

The comparison between the pulse height of simulated pulses and real pulses induced by the on-chip heater as a function of the energy is shown in Fig. 15. There is a good agreement between the datasets over the whole detector energy bandwidth, up to the saturation regime. Saturation occurs at $E_{SAT} \sim 4$ MeV, which corresponds to the raw expected value. Given the assessed heat capacities at the TES operating point and the width of the transition from the TES operating point to the its saturation ($T_{SAT} \sim 112$ mK), we indeed have:

$$E_{SAT} \sim C_{TOT,0} \cdot \Delta T_{SAT}$$
$$\sim (C_{TES,0} + C_{ABS,0}) \cdot (T_{SAT} - T_{TES,0})$$
$$\sim 62\,pJ/K \cdot 10\,mK \sim 4\,MeV \tag{15}$$

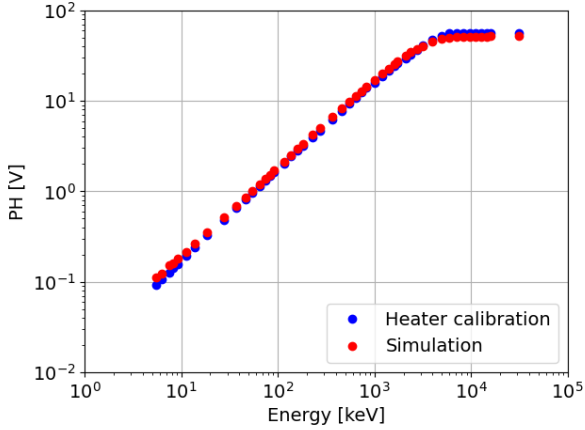

**Fig. 15.** Pulse Height vs deposited Energy: comparison between simulator output (red points) and DM127 data (blue points).

### C. Different working points

In the previous section, we have compared the simulator output with real data obtained by operating the DM127 at its optimal operating point (see Fig. 16 for a summary). This is the same point at which we carried out the assessment of the detector heat capacities, by finding their optimal correction factors. To obtain a more independent validation, without having another detector available, we have performed the same comparison between simulator output and real data by operating the detector in a completely different operating point. Details of the two operating conditions are given in Tab. 6.

**Tab. 6.** - Comparison between the different DM127 operating points used to validate the simulator.

| Parameter | Optimal operating point | Second operating point |
|---|---|---|
| $I_{bias}$ | 1000 µA | 600 µA |
| $R_0$ | 139 mΩ | 95 mΩ |
| $T_{TES,0}$ | 102 mK | 90 mK |
| $\alpha_0$ | 50 | 10 |

Fig. 17 - Left shows the new detector operating point, and Fig. 17 - Right the comparison between a simulated 150 keV pulse and real data from cosmic muon events. The agreement between real and simulated data is slightly worse with respect to the optimal operating point, but still acceptable for the purposes of the CryoAC simulation (~ 10% in pulse height). This could be related to the rough assessment of the detector heat capacities, but also to the simplified TES transition description used in the simulator (red lines in Fig. 16-17), which does not take into account local transition features.

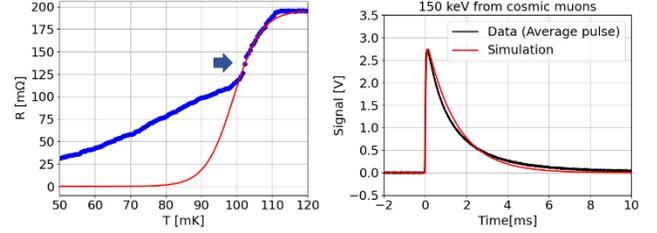

**Fig.16.** Simulator validation at the optimal operating point of the DM127. (Left) The optimal operating point marked by an arrow on the TES transition ($R_0$ = 139 mΩ). The red line shows the transition description in the simulator. (Right) Comparison between simulator output and DM127 data, referring to 150 keV energy deposition from a cosmic muon.

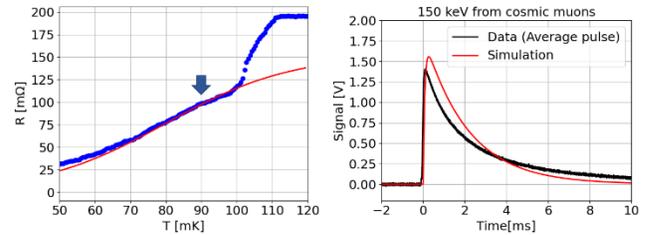

**Fig. 17.** Simulator validation in a second operating point of DM127. (Left) The second operating point marked by an arrow on the TES transition ($R_0$ = 90 mΩ). The red line shows the transition description in the simulator. (Right) Comparison between simulator output and DM127 data, referring to 150 keV energy deposition from a cosmic muon.

## VI. CONCLUSION

We are developing the ATHENA X-IFU Cryogenic Anticoincidence Detector (CryoAC). The detector baseline has recently changed from an athermal design to a more classical microcalorimeter thermal scheme, which will be used for the next Demonstration Model prototype, i.e. the CryoAC DM1.1.

In this context, detector development is now being driven by a new thermal end-to-end simulator. We have used the latest CryoAC single-pixel prototype to measure the thermal parameters of the detector, which have been inserted into the simulator. It has been then validated by comparing its output with real data.

The simulator is capable of describing pulse shape, pulse height and detector noise with an accuracy of a few percent over a wide energy range (from a few keV up to the saturation regime above ~ 4 MeV). Furthermore, it is robust in very different detector operating conditions.

We consider this accuracy to be fully sufficient for the aims of the simulator, which will be used to finalize the CryoAC cold stage design and conduct its performance assessment.

## REFERENCES

[1] X. Barcons et al., "Athena: ESA's X-ray observatory for the late 2020s", Astron. Nachr./ AN 338:153–158 (2017), https://doi.org/10.1002/asna.201713323.

[2] P. Peille et al., "The X-ray Integral Field Unit at the end of the Athena reformulation phase," submitted in 2024 to Springer Nature.

[3] D. Barret et al., "The Athena X-ray Integral Field Unit: a consolidated design for the system requirement review of the preliminary definition phase," Exp. Astr., vol. 55, pp. 373-476, Jan. 2023, https://doi.org/10.1007/s10686-022-09880-7.

[4] C. Macculi et al., "The Cryogenic Anticoincidence detector for the new Athena X-IFU instrument: a program overview", Condens. Matter 2023, 8, 108 2023), https://doi.org/10.3390/condmat8040108

[5] S. Lotti et al., " Review of the Particle Background of the Athena X-IFU Instrument", ApJ 909 111 (2021), https://doi.org/10.3847/1538-4357/abd94c

[6] L. Ferrari Barusso et al., "First Configurational Study of the CryoAC Detector Silicon Chip of the Athena X-Ray Observatory," IEEE Trans. Appl. Supercond., vol. 33, no. 1, pp. 1-7, Jan. 2023, https://doi.org/10.1109/TASC.2022.3222959.

[7] M. D'Andrea et al., "The end-to-end simulator of the ATHENA X-IFU Cryogenic AntiCoincidence detector (CryoAC)", Proceedings Volume 13093, Space Telescopes and Instrumentation 2024: Ultraviolet to Gamma Ray; 130930X (2024) https://doi.org/10.1117/12.3017794

[8] L. Ferrari Barusso et al. "Status of the Cryogenic Anti-Coincidence Detector (CryoAC) for the Athena X-ray Integral Field Unit (X-IFU)", IEEE Trans. Appl. Supercond., This Special Issue (2024) https://doi.org/10.1109/TASC.2024.3518459

[9] M. Kiviranta et al., "Two-stage SQUID amplifier for the frequency multiplexed readout of the X-IFU X-ray camera," IEEE Trans. Appl. Supercond., vol. 31, no. 5, pp. 1-5, Aug. 2021, https://doi.org/10.1109/TASC.2021.3060356.

[10] M. D'Andrea et al., "The TES-based Cryogenic AntiCoincidence Detector (CryoAC) of ATHENA X-IFU: A Large Area Silicon Microcalorimeter for Background Particles Detection," JLTP, vol. 214, pp. 164-172, Jan. 2023, https://doi.org/10.1007/s10909-023-03034-5.

[11] K. Irwin, G. Hilton, "Transition-edge sensors", in Cryogenic Particle Detection. Topics in Applied Physics, vol. 99, ed. by C. Enss (2005), https://doi.org/10.1007/10933596_3..

[12] M. D'Andrea et al., "The Demonstration Model of the ATHENA X-IFU Cryogenic AntiCoincidence Detector," JLTP, vol. 199, pp. 65-72, Jan. 2020, https://doi.org/10.1007/s10909-019-02300-9.

[13] D. Bagliani et al., "Ir TES Electron-Phonon Thermal Conductance and Single Photon Detection", JLTP, vol. 151, pp. 234-238, (2008), https://doi.org/10.1007/s10909-007-9641-1.

[14] M. D'Andrea, "Opening the low-background and high-spectral-resolution domain with the ATHENA large X-ray observatory: Development of the Cryogenic AntiCoincidence", PhD thesis (2019), https://arxiv.org/abs/1904.03307.